\documentclass[12pt]{iopart}

\usepackage{graphicx}
\usepackage[usenames]{color}
\newcommand{\beq}{\begin{equation}}
\newcommand{\eeq}{\end{equation}}
\newcommand{\beqa}{\begin{eqnarray}}
\newcommand{\eeqa}{\end{eqnarray}}
\newcommand{\ba}{\begin{array}}
\newcommand{\ea}{\end{array}}

\usepackage{iopams}

\begin{document}

\title[Bright solitons in a quasi-1D reduced model of a dipolar BEC]{Bright solitons in a quasi-one-dimensional reduced model of a
dipolar Bose-Einstein condensate with repulsive short-range interactions}

\author{Emerson Chiquillo}

\address{Instituto de F\'{\i}sica Te\'orica, UNESP - Universidade Estadual Paulista,
\\ 01.140-070 S\~ao Paulo, S\~ao Paulo, Brazil}
\ead{emerson.chiquillo@gmail.com}

\begin{abstract}

We study the formation and dynamics of bright solitons in a quasi-one-dimensional 
reduced mean-field Gross-Pitaevskii equation of a dipolar Bose-Einstein condensate
with repulsive short-range interactions.
The study is carried out using a variational approximation and a numerical solution.
Plots of chemical potential and root mean square (rms) size of solitons are obtained 
for the quasi-one-dimensional model of three different dipolar condensates of
 $^{52}$Cr, $^{168}$Er and $^{164}$Dy atoms. 
The results achieved are in good agreement
with those produced by the full three-dimensional mean-field model of the 
condensate.
We also study the dynamics of the collision of a train of two solitons in the
quasi-one-dimensional model of every condensate above.
At small velocities (zero or close to zero) the dynamics is attractive for a phase difference $\delta=0$,
the solitons coalesce and these oscillate forming a bound soliton molecule. 
For a phase difference $\delta=\pi$ the effect is repulsive. 
At large velocities the collision is 
independent of the initial phase difference $\delta$.
This is quasi-elastic and the result is two quasi-solitons.

\end{abstract}


\maketitle

\section{Introduction}

A stable bright soliton is a self-focusing nonlinear wave arises from balance
between dispersive effects and nonlinear attractive interactions. This solitary wave 
maintains its shape while traveling at constant velocity. 
In a Bose-Einstein condensate (BEC) under harmonic transverse confinement 
the bright solitons appear for attractive short-range interactions
\cite{solit-1}. 
The formation of quasi-one-dimensional (quasi-1D) bright solitons and
bright soliton trains was observed experimentally in BECs of $^{7}$Li 
\cite{solit-Li-1,solit-Li-2} and $^{85}$Rb \cite{solit-Rb} atoms
by turning the repulsive interatomic interaction until an attractive interaction
via Feshbach resonances (FR) and releasing the condensate 
in an expulsive trap or an axially free. 

The experimental realization of BECs with interatomic magnetic 
dipole dipole interactions or dipolar BECs (DBECs) in $^{52}$Cr \cite{Cr-1,Cr-2,Cr-3,Cr-4,Cr-5,Cr-6},
 $^{164}$Dy \cite{Dy-1} and $^{168}$Er \cite{Er-1} atoms
has opened the door to a new level in the research of solitons 
in degenerate quantum gases. This is because of the interesting features related
to the anisotropic long-range character of the dipolar interaction.
Experimentally polar molecules with large electric dipole moments
were also used \cite{Electric}.

Under appropriate conditions the existence of bright solitons may be 
regarded as a consequence of the anisotropy associated with the 
dipole dipole interaction.
In \cite{Two Dim Bright} for a disk-shaped dipolar condensate the necessary conditions 
for the existence of stable 2D bright solitons trapped in the axial direction 
by reversing the sign of the dipolar interaction
(by rotating the magnetic field for magnetic dipoles) and the
dipoles aligned in the same direction of the trap are discussed. 
For a quasi-2D dipolar condensate with positive dipolar strength and 
atoms with the dipole moments polarized perpendicular
to direction of the trap, stable anisotropic bright solitons were predicted  along with the
respective requirements to the collapse \cite{Anisotropic Solitons}. 
The collision of quasi-2D anisotropic bright solitons in a DBEC was performed in 
\cite{Anisotropic 2D sollision}. The analysis was carried out using both a 
time-dependent variational approximation  and a full numerical solution. 
\cite{Experimental anisotropic 2D-solitons} shows the first experimental realization 
of a quasi-2D anisotropic bright soliton in a dipolar condensate. 
Recently, was proposed the existence of robust 1D and 2D bright solitons 
in a BEC with repulsive dipolar interactions induced by a combination of polarizing fields,
oriented perpendicular to the plane in which the condensate is trapped \cite{Solitons induced}.
In a quasi-1D DBEC solitons exist for an arbitrarily large number of atoms \cite{Cuevas}
because of there is no collapse in one-dimensional models with cubic nonlinearity. 
In a 3D DBEC considering positive dipolar strength and repulsive atomic interaction, 
the numerical solution of the full three-dimensional mean-field Gross-Pitaevskii equation
(GPE) with confinement in the radial direction and free in the axial one, 
supports bright solitons below a critical number of 
atoms when the scattering length is less than the dipolar strength \cite{Luis Y}.
So beyond a critical value of atom number the condensate is unstable and it should collapse.
We show the existence of bright solitons for repulsive contact interaction in a DBEC 
using a quasi-1D reduced model of the GPE \cite{Dim-reduc}. We also study the dynamics
of these solitons. 
The results of the quasi-1D equation are in good agreement with those produced by the full
3D condensate. Our findings show that the solitons exist provided the value 
of the scattering length is
less than the dipolar strength and there is no collapse in this reduced 1D model.
We also found that the collision between two bright solitons
is sensitive to the initial phase difference 
at low velocities (close to zero or zero). At high velocities the collision is quasi-elastic and it is 
independent of the initial phase difference. In \sref{sec2} we present the 1D reduced mean-field model for studying the statics and dynamics of a   
dipolar BEC without confinement trap. We also include a Gaussian variational approximation 
of this equation. The numerical and variational results are shown in \sref{sec3}. 
Finally, we present a brief summary and discussion of our study in \sref{sec4}.

\section{Analytical Consideration}\label{sec2}
\subsection{1D reduced mean-field model for a dipolar condensate}

In a trapped dipolar condensate many static and dynamic properties can be described 
using the time-dependent non-local 3D GPE. 
The numerical solution of this equation is a complicated
issue because of the anisotropic long-range character of the dipolar interaction.
However, in many experimental arrangements with extreme symmetry of the 
harmonic traps the dynamics of the condensate takes place in reduced dimensions
and the system becomes disk- or cigar-shaped.
A quasi-1D model of the GPE describes a cigar-shaped condensate. 
The numerical solution and the variational approximation of such quasi-1D equation
is simpler than that of the full 3D equation \cite{Dim-reduc}. 

We study bright solitons in a dipolar BEC of $N$ atoms, each of mass $m$ by means of the
dimensionless dipolar GPE (DGPE). We use as unit of length the oscillator length 
$l_{\perp}=\sqrt{\hbar/m\omega_{\perp}}=1\,\mu m$. The time $t$ is measured in units of
$\omega_{\perp}^{-1}$ and the energy in units $\hbar\omega_{\perp}$.
So the mean-field DGPE is \cite{Cr-2,Cr-5,Cr-6,Luis Y, Dim-reduc}
\begin{eqnarray}\label{eq1-dip}
\fl {\mbox i} \frac{\partial \Psi({\bf r},t)}{\partial t}
=  {\Biggl [}  -\frac{1}{2}\nabla^2
+ \frac{1}{2}\rho^2 +4\pi aN\vert \Psi({\bf r},t)\vert^2\nonumber \\[0.3true cm]
+ 3Na_{dd}\int \rmd\mathbf{r}' 
\left(\frac{1-3\cos^{2}\theta}{\left|\mathbf{r}-\mathbf{r}'\right|^{3}}\right)
\vert \Psi({\bf r'},t)\vert^2 {\Biggl ]} \Psi({\bf r},t)
\end{eqnarray}
where $\Psi({\bf r},t)$ is the condensate wave function  with normalization 
$ \int d\mathbf{r}\left|\Psi\left(\mathbf{r},t\right)\right|^{2}=1 $.
$\rho^2/2$ is the radial harmonic trap, with $\rho\equiv(x,\,y)=\sqrt{x^{2}+y^{2}}$.
The atomic scattering length is $a$. The constant
$a_{dd}=\mu_0 \widetilde {\mu}^2m/12\pi\hbar^2$ is 
the strength of the dipolar interaction with $\mu_0$ the permeability of free space and 
$\widetilde{\mu}$ the magnetic dipole moment of a single atom. $\theta$ is the angle between 
the relative position of the dipoles $ \left|\mathbf{r}-\mathbf{r}'\right|$ 
and the polarization direction $z$. 

The wave function propagates in time, e.g. as
$\Psi\left(\mathbf{r},t\right)=\Psi\left(\mathbf{r}\right)\exp\left(-\rmi \mu t\right)$
\cite{libro Pethick, art Dalfovo}, where $\mu$ is the chemical potential. 
Using this ansatz in the equation (\ref{eq1-dip}) we get the time-independent dipolar GPE 
\begin{eqnarray}\label{time indep DGPE}
\fl \mu\Psi\left(\mathbf{r}\right)
= {\Biggl [}  -\frac{1}{2}\nabla^2
+ \frac{1}{2}\rho^2 +4\pi aN\vert \Psi({\bf r})\vert^2\nonumber \\ [0.3true cm]
+ 3Na_{dd}\int \rmd\mathbf{r}' 
\left(\frac{1-3\cos^{2}\theta}{\left|\mathbf{r}-\mathbf{r}'\right|^{3}}\right)
\vert \Psi({\bf r'})\vert^2 {\Biggl ]} \Psi({\bf r})
\end{eqnarray}
From the normalization condition of the wave function and 
the thermodynamic relation $\mu=\partial E/\partial N$,
the energy of the condensate is given by
\begin{eqnarray}\label{dipolar energy}
\fl E\left(\Psi\right)
= \int \rmd\mathbf{r} {\Biggl [} \frac{N}{2}\left|\nabla\Psi\left(\mathbf{r}\right)\right|^{2}
+ \frac{N}{2}\rho^2\left|\Psi\left(\mathbf{r}\right)\right|^{2}
+2\pi aN^{2}\left|\Psi\left(\mathbf{r}\right)\right|^{4} \nonumber \\ [0.3true cm]
+ \frac{3a_{dd}N^{2}}{2} \int \rmd\mathbf{r}' 
\left(\frac{1-3\cos^{2}\theta}{\left|\mathbf{r}-\mathbf{r}'\right|^{3}}\right)
\vert \Psi({\bf r'})\vert^2 \vert \Psi({\bf r})\vert^2 {\Biggl ]} 
\end{eqnarray}

For a cigar-shaped DBEC the reduction from 3 to 1 dimension is achieved 
using the adiabatic approximation \cite{Effective wave equations,
Effective mean-field equations for,reduction}, where we consider that 
in a cigar-shaped trap with tight radial binding the dynamics takes place
along the axial direction and in the transverse direction the condensate wave function
remains in the ground state. So the wave function to 
the mean-field equation (\ref{eq1-dip}) can be decomposed as 
$\Psi\left(\mathbf{r},t\right)=\kappa\left(\rho\right)\phi\left(z,t\right)$, with
\begin{eqnarray}
\Psi\left(\mathbf{r},t\right)=\left(\frac{1}{\pi a_{\rho}^{2}}\right)^{1/2}
\exp\left(-\frac{\rho^{2}}{2\, a_{\rho}^{2}}\right)\phi\left(z,t\right)
\end{eqnarray}
where $a_{\rho}$ is the radial harmonic oscillator length and it fulfill
$\omega_{\perp}a_{\rho}^{2}=1$. 

We get the quasi-one-dimensional DGPE 
by substituting this ansatz in the DGPE (\ref{eq1-dip}), 
multiplying by $\kappa^{*}\left(\rho\right)$
and integrating over $\rho$ (using the normalization
$\int \rmd ^{2}\rho\left|\kappa\right|^{2}=1$). 
The quasi-1D dipolar contribution can be obtained easily using the Fourier
transform of the 3D dipole dipole energy performing the dimensional reduction in the
momentum space and returning to the configuration space \cite{Dim-reduc}.
So the quasi-1D DGPE is given by
\begin{eqnarray}\label{eq2-red-dip}
\fl {\mbox i}\frac{\partial\phi\left(z,t\right)}{\partial t}
= \biggl[-\frac{1}{2}\frac{\partial^{2}}{\partial z^{2}}+
\frac{2\, aN}{a_{\rho}^{2}}\left|\phi\left(z,t\right)\right|^{2}
 \nonumber \\ [0.3true cm] 
 +N\int_{-\infty}^{\infty}\rmd z'\, U_{dd}^{1D}\left(z-z'\right)
\left|\phi\left(z',t\right)\right|^{2}\biggr]\phi\left(z,t\right)
\end{eqnarray}
where the 1D potential in configuration space is
\begin{eqnarray}\label{eq-energy dipolar transf-2}
U_{dd}^{1D}\left(z-z'\right)
=\frac{1}{2\pi}\int_{-\infty}^{\infty}\rmd q_{z}\exp\left(iq_{z}z\right)V_{1D}\left(q_{z}\right)
\end{eqnarray}
and the 1D potential in Fourier space is
\begin{eqnarray}
V_{1D}\left(q_{z}\right)=
2a_{dd}\int_{0}^{\infty}\rmd q_{\rho}\, 
q_{\rho}\left(\frac{3q_{z}^{2}}{q_{\rho}^{2}+q_{z}^{2}}-1\right)
\exp\left(-\frac{q_{\rho}^{2}\, a_{\rho}^{2}}{2}\right)
\end{eqnarray}
\subsection{Variational approximation}

An understanding of the existence of bright solitons in the cigar-shaped DBEC 
can be obtained from a variational approximation to the equation (\ref{eq2-red-dip})
with the following time-independent Gaussian ansatz
\begin{eqnarray}\label{eq-Gauss}
\phi\left(z\right)= 
\left(\frac{1}{w_{z}^{2}\pi}\right)^{1/4}
\exp\left(-\frac{z^{2}}{2w_{z}^{2}}\right)
\end{eqnarray}
where $w_{z}$ is the variational width along the $z$ direction. However, calculation of energy 
using this wave function in the quasi-1D model (\ref{eq2-red-dip})
involves integrals of the non-trivial dipolar contribution (\ref{eq-energy dipolar transf-2}).
Instead, it is easier to obtain the 3D Gaussian energy and considering 
the quasi-1D energy as a special case of this. Thus, using the conventional
3D Gaussian ansatz \cite{Cr-1,Luis Y,Dim-reduc}
\begin{eqnarray}
\Psi\left(\rho,z\right)=
\left(\frac{1}{w_{\rho}^{2}\, w_{z}\pi^{3/2}}\right)^{1/2}
\exp\left[-\left(\frac{\rho^{2}}{2w_{\rho}^{2}}+\frac{z^{2}}{2w_{z}^{2}}\right)\right]
\end{eqnarray}
where $w_{\rho}$ and $w_{z}$ are the variational widths along the radial
$\rho$ and axial $z$ directions, respectively. 
By substituting this wave function in the 3D dipolar energy  (\ref{dipolar energy}),
we get 
\begin{eqnarray}
\fl E 
= \frac{N}{4}\biggl[2\left(1+\frac{1}{w_{\rho}^{4}}\right)w_{\rho}^{2}
+\frac{1}{w_{z}^{2}}\biggl] \nonumber \\ [0.3true cm] 
+\frac{N^{2}}{\sqrt{2\pi}w_{\rho}^{2}\, w_{z}} \big[a-a_{dd}f\left(\kappa\right)\big]
\end{eqnarray}
where
$f(\kappa)=\big[1+2\kappa^{2}-3\kappa^{2}d\left(\kappa\right)\big]/\big({1-\kappa^{2}}\big)$
with
$d\left(\kappa\right)=\mathrm{tanh^{-1}}\sqrt{1-\kappa^{2}}/\sqrt{1-\kappa^{2}}$
and 
$\kappa=w_{\rho}/w_{z}$.
The function $f(\kappa)$ \cite{Cr-1,Two Dim Bright,97 Exact hydrodynamics dbec,funcion f(k)} 
has asymptotic values $f\left(\kappa\rightarrow0\right)=1$
(cigar-shaped DBEC) and $f\left(\kappa\rightarrow\infty\right)=-2$
(disk-shaped DBEC) and it changes the sign at $\kappa=1$.
The contribution of the function $f(\kappa)$ also can be seen 
through the variational treatment in \cite{Anisotropic Solitons} for anisotropic 2D 
bright solitons. The asymptotic behaviour of the Equation 6 of \cite{Anisotropic Solitons} 
in the isotropic case $\kappa_{x}=\kappa_{y}$ corresponds with our function $f(\kappa)$ such that 
$f(\kappa)\equiv -h(\kappa_{x}=\kappa_{y})$.

The energy in the quasi-1D model $E_{1D}$ can be obtained using  the radial harmonic oscillator length
$a_{\rho}$, such that $w_{\rho}=a_{\rho}$
in the dipolar contribution and neglecting the derivatives respect to $w_{\rho}$
in the kinetic energy term \cite{Dim-reduc}. These considerations lead to
\begin{eqnarray}\label{Dipolar energy 1D}
E_{1D} = \frac{N}{4w_{z}^{2}}
+\frac{N^{2}}{\sqrt{2\pi}a_{\rho}^{2}\, w_{z}}\left[a-a_{dd}f\left(\kappa_{0}\right)\right]
\end{eqnarray}
where $\kappa_{0}=a_{\rho}/w_{z}$. In a extreme quasi-1D DBEC, the axial width is much 
larger than the transverse oscillator length, as well as $\kappa_{0}\rightarrow0$
then $f\left(\kappa_{0}\right)\rightarrow1$ and the dipolar energy becomes 
$-N^{2}a_{dd}/\left(\sqrt{2\pi}a_{\rho}^{2}\, w_{z}\right)$. 
Therefore the variational approximation indicates 
that the dipolar interaction turns into a contact interaction.
So the total interaction (contact and dipolar) is an effective contact interaction such that
\begin{eqnarray}
a_{eff}=a-a_{dd}
\end{eqnarray}
This effective scattering length becomes attractive when $a_{dd}>a$, i.e. it is possible 
to have bright solitons even for repulsive scattering length $(a>0)$.
This interesting scenario was shown for the first time in a 3D dipolar condensate \cite{Luis Y}.
The bright solitons can be regarded as stable stationary states and these are given by the
minimum of the energy. So by a minimization of 1D ground state energy
(\ref{Dipolar energy 1D}) by $\partial E_{1D}/\partial w_{z}=0$ the equation to the width
of the condensate $w_z$ \cite{Dim-reduc} is
\begin{eqnarray}
\frac{1}{2w_{z}^{3}}+
\frac{N\left[a-a_{dd}\, h\left(\kappa_{0}\right)\right]}{\sqrt{2\pi}\, a_{\rho}^{2}\, w_{z}^{2}}=0
\end{eqnarray}
where the function $h\left(\kappa_0\right)$ can be written as
\begin{eqnarray}
h(\kappa)=
\frac{1+10\kappa_0^{2}-2\kappa_0^{4}-9\kappa_0^{2}d(\kappa_0)}
{(1-\kappa_0)^{2}}
\end{eqnarray}

\section{Numerical and variational results}\label{sec3}

\begin{figure}
\begin{center}
\includegraphics[width=.49\linewidth]{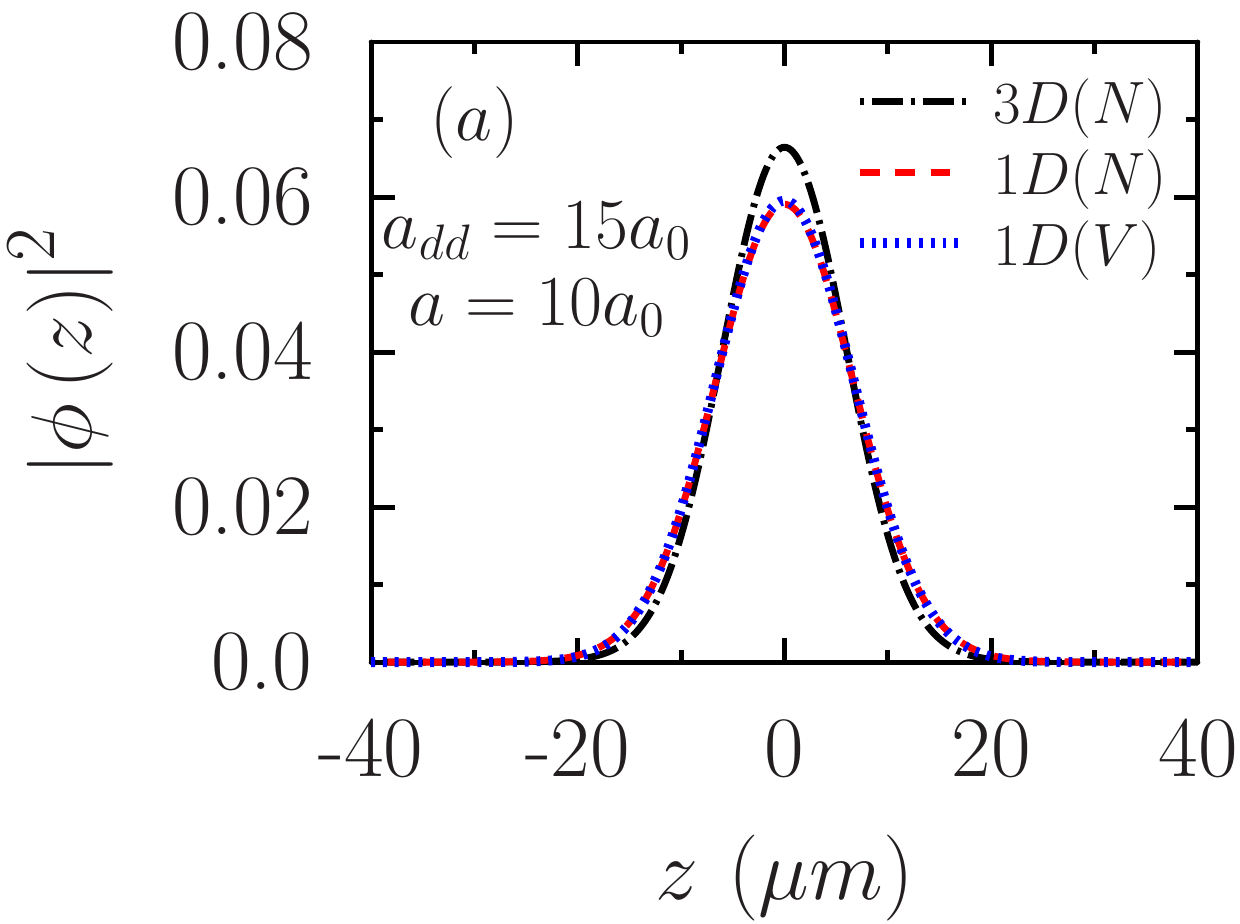}
\includegraphics[width=.49\linewidth]{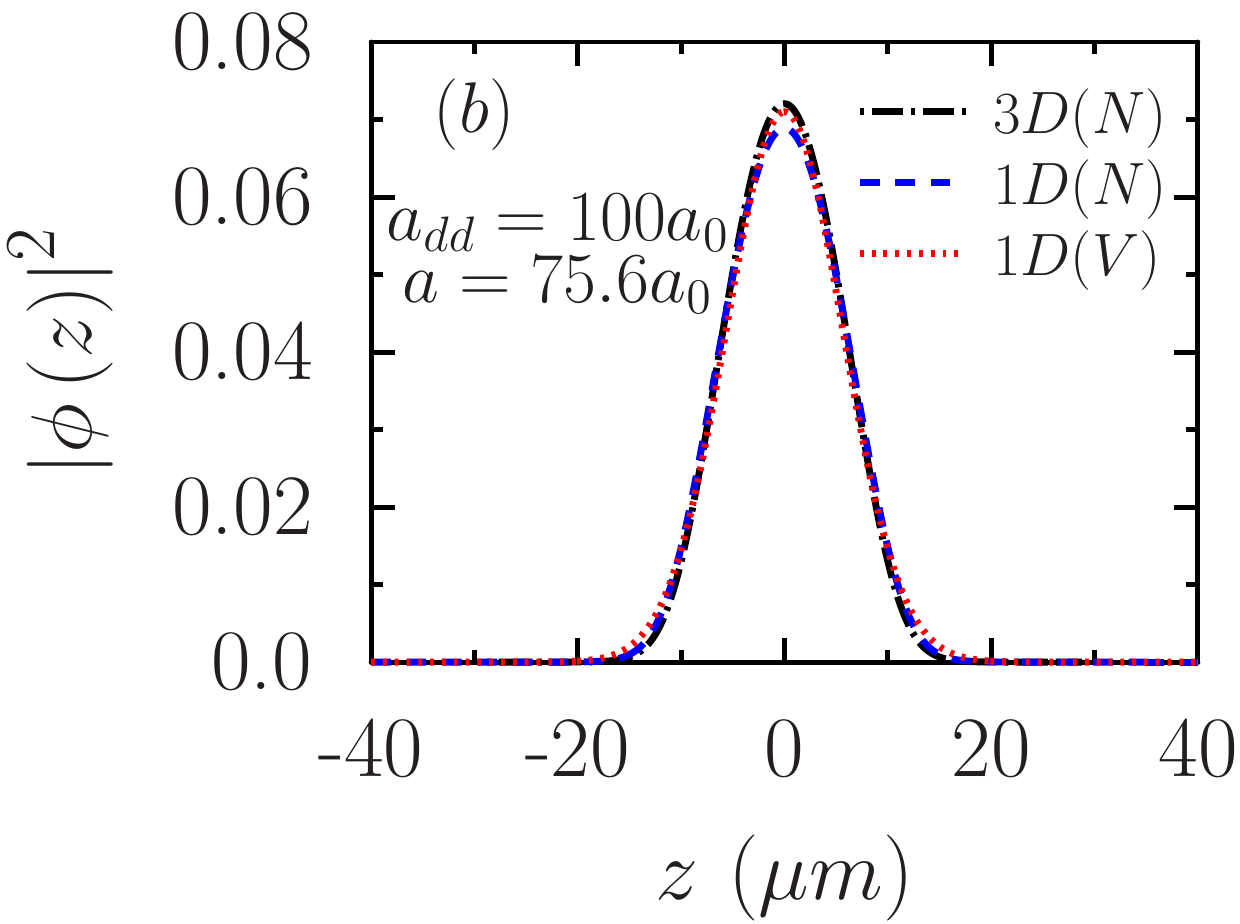}
\end{center}

\caption{ Profile of $\left|\phi\left(z\right)\right|^{2}$
of a dipolar BEC of 1000 atoms. 
Numerical solution (N) of 3D (\ref{eq1-dip})  and quasi-1D model (\ref{eq2-red-dip}). 
The variational approximation (V) for the quasi-1D model is given by (\ref{eq-Gauss}).
$\left(a\right)$  $ a_{dd}=15a_{0}$ and $a=10a_{0}$. $\left(b\right)$  $a_{dd}=100a_{0}$ 
and $a=75.6a_{0}$. The oscillator length used is $1\,\mu m$.}
\label{fig1}
\end{figure}

The 3D and quasi-1D GPEs are solved numerically by the split-step Crank-Nicolson method in 
Cartesian coordinates \cite{Adhikari comput}.
The 3D dipolar contribution is evaluated in the momentum space using the convolution theorem as 
\begin{eqnarray}
\fl \int d\mathbf{r}'\, U_{dd}\left(\mathbf{\mathbf{r}-\mathbf{r}'}\right)
\left|\Psi\left(\mathbf{r'}\right)\right|^{2}=
\nonumber\\ [0.25true cm] 
\mathcal{F}^{\,-1}\left[\mathcal{F}\left\{ U_{dd}\right\} 
\left(\mathbf{\mathbf{q}}\right)\mathcal{F}
\left\{ \left|\Psi\right|^{2}\right\}
\left(\mathbf{\mathbf{q}}\right)\right]\left(\mathbf{r}\right)
\end{eqnarray}
with $U_{dd}(\mathbf{r}-\mathbf{r}')=
3a_{dd}\big(1-3\cos^{2}\theta\big)/\left|\mathbf{r}-\mathbf{r}'\right|^{3}$. 
The Fourier transform of the 3D dipolar potential 
$\mathcal{F}\left\{ U_{dd}\right\} \left(\mathbf{\mathbf{q}}\right)$
is given analytically by \cite{Cr-1,Mean-field regime of trapped DBEC,Trans Fourier}
\begin{eqnarray}
\mathcal{F}\left\{ U_{dd}\right\} \left(\mathbf{\mathbf{q}}\right)=
4\pi a_{dd}\left(3\frac{q_{z}^{2}}{\mathbf{q}^{2}}-1\right)
\end{eqnarray}
The term $\mathcal{F}\left\{ \left|\Psi\right|^{2}\right\} \left(\mathbf{\mathbf{q}}\right)$
is evaluated numerically using the fast Fourier transform (FFT) in Cartesian 
coordinates  \cite{Fast FT}. The inverse Fourier transform is obtained by means of the FFT.
The quasi-1D dipolar contribution (\ref{eq-energy dipolar transf-2}) is also calculated using FFT.

\begin{figure}
\begin{center}
\includegraphics[width=.6\linewidth]{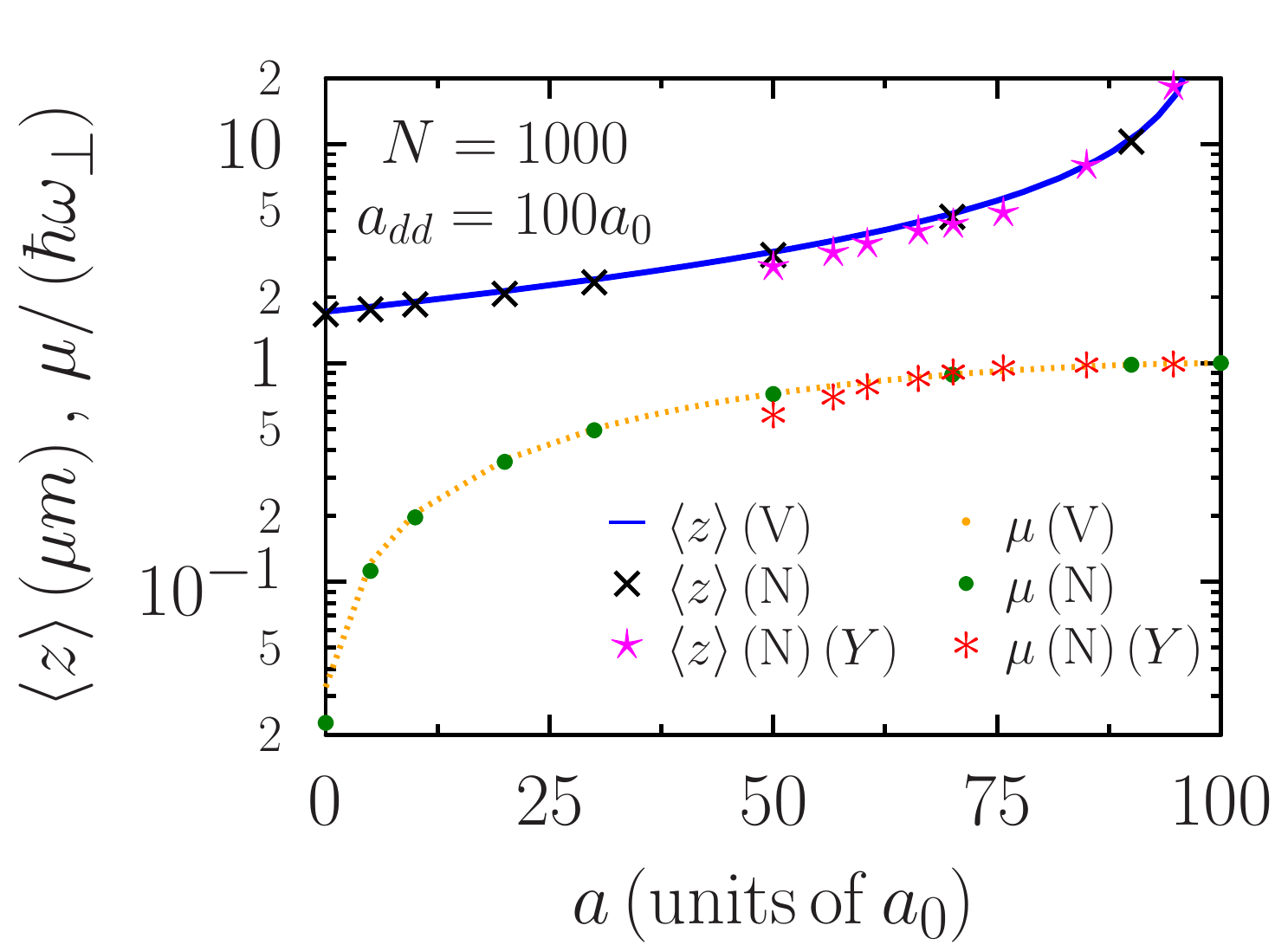}
\end{center}

\caption{Chemical potential $\mu$ and rms size $\left\langle z\right\rangle$ versus 
the scattering length $a$ of a quasi-1D and a 3D dipolar condensate of 1000 atoms
with dipolar strength $a_{dd}=100a_{0}$. Numerical solution (N), variational approximation (V)
and Young {\it et al.} (Y) \cite{Luis Y}. The oscillator length used is 
$1\,\mu m $.}
\label{fig2}

\end{figure}

In Figures \ref{fig1}$a$ and $1b$ 
we plot the numerical and variational results 
of the profile of $\left|\phi\left(z\right)\right|^{2}$ 
from quasi-1D reduced mean-field  DGPE (\ref{eq2-red-dip}) and the 
Gaussian ansatz (\ref{eq-Gauss}) respectively.
We also shown the numerical results of the full 3D equation (\ref{eq1-dip}). 
We use two condensates each of 1000 atoms
with dipolar strengths $a_{dd}=15a_{0}$ ($a=10a_{0}$) and 
$a_{dd}=100a_{0}$ ($a=75.6a_{0}$), where $a_0$ is the Bohr radius.
In both cases, variational and numerical results of the quasi-1D equation 
are in good agreement with those of the full 3D model.
In the Figure \ref{fig2} 
we compare numerical and variational results of
chemical potential $\mu$ and root mean square (rms) $\left\langle z\right\rangle$
as functions of the scattering length using the quasi-1D model 
for a dipolar strength $a_{dd}=100a_{0}$ 
with the numerical results of full 3D equation obtained from \cite{Luis Y}.

\begin{figure}
\begin{center}

\includegraphics[width=.49\linewidth]{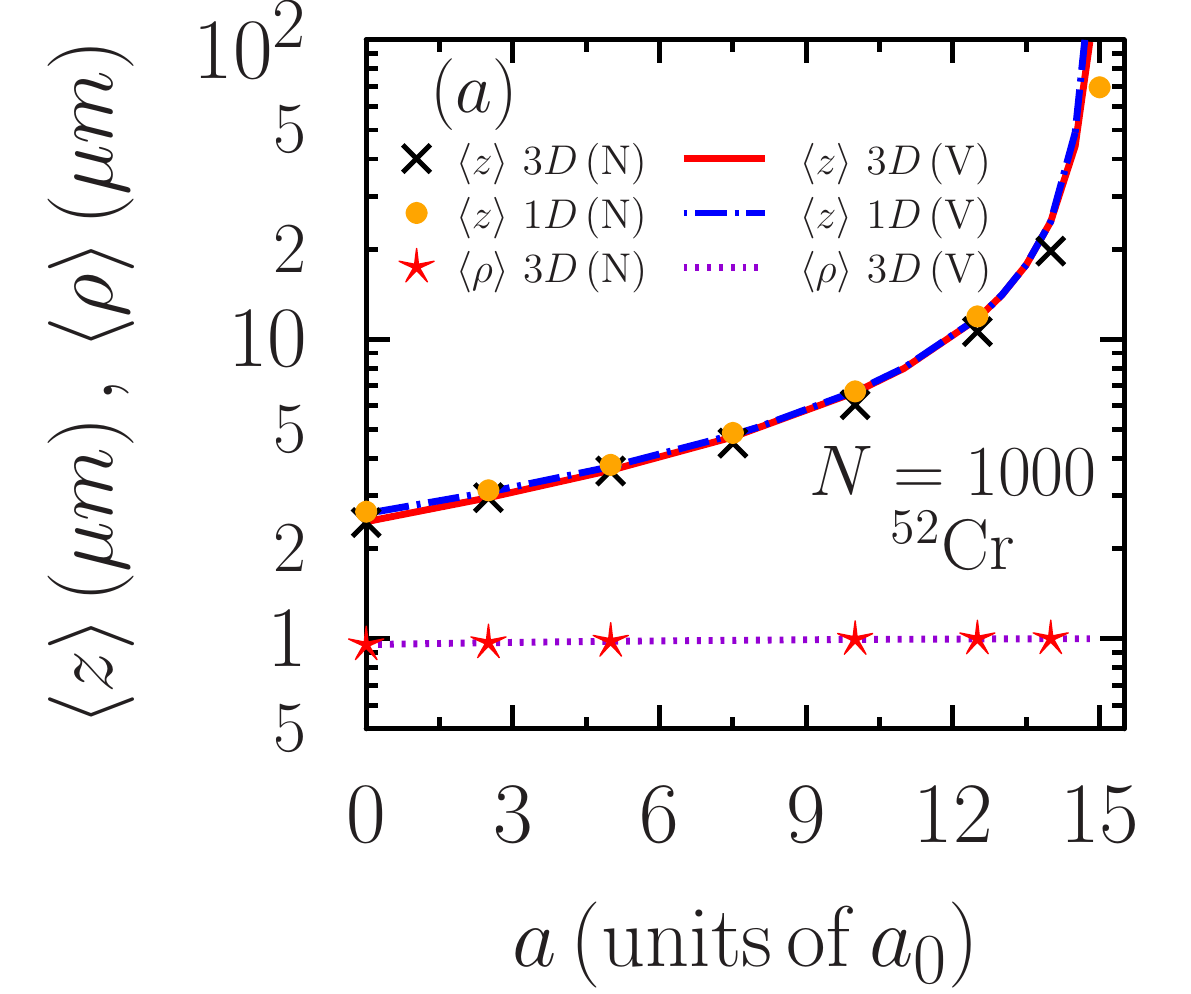}
\includegraphics[width=.49\linewidth]{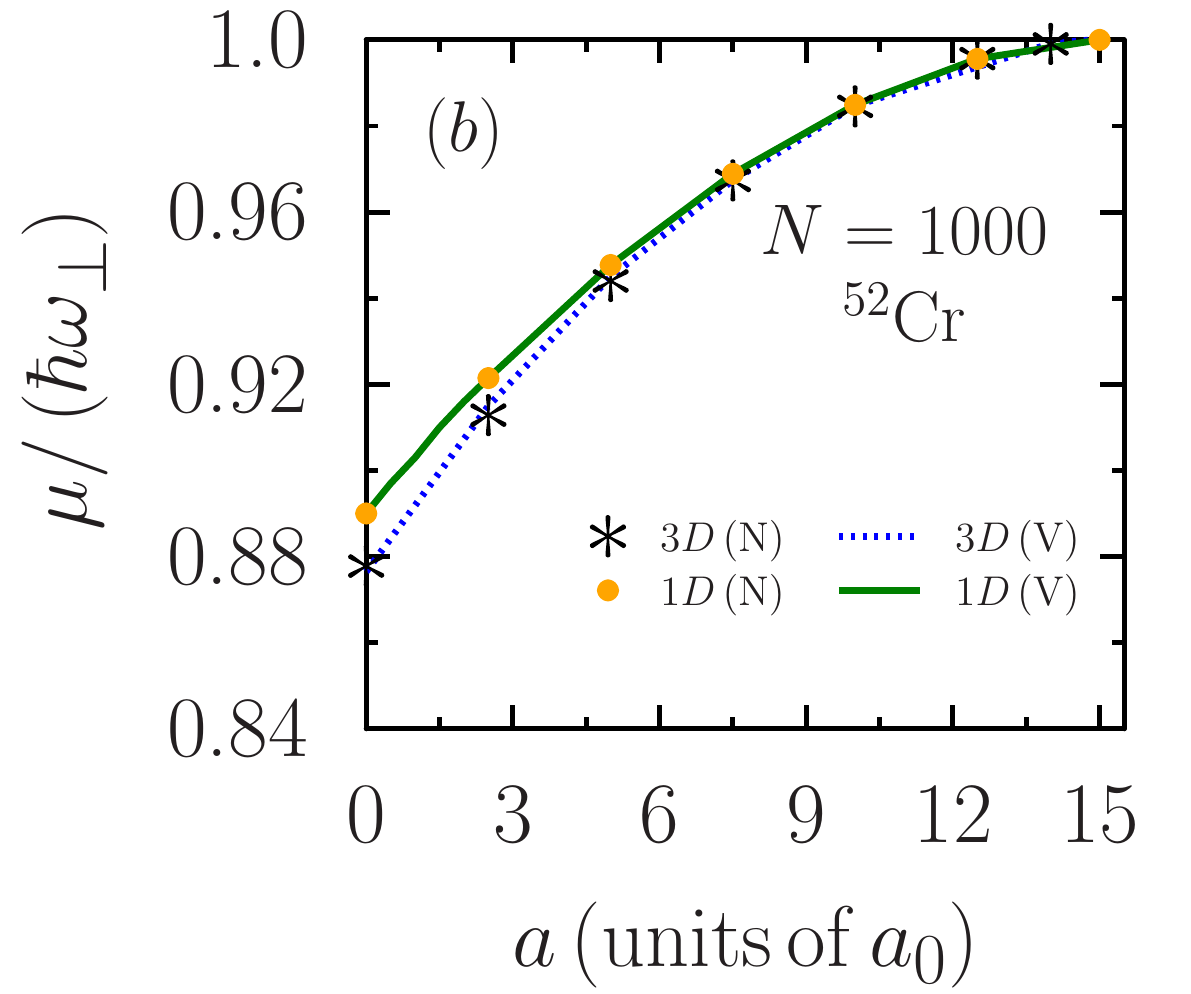}
\includegraphics[width=.49\linewidth]{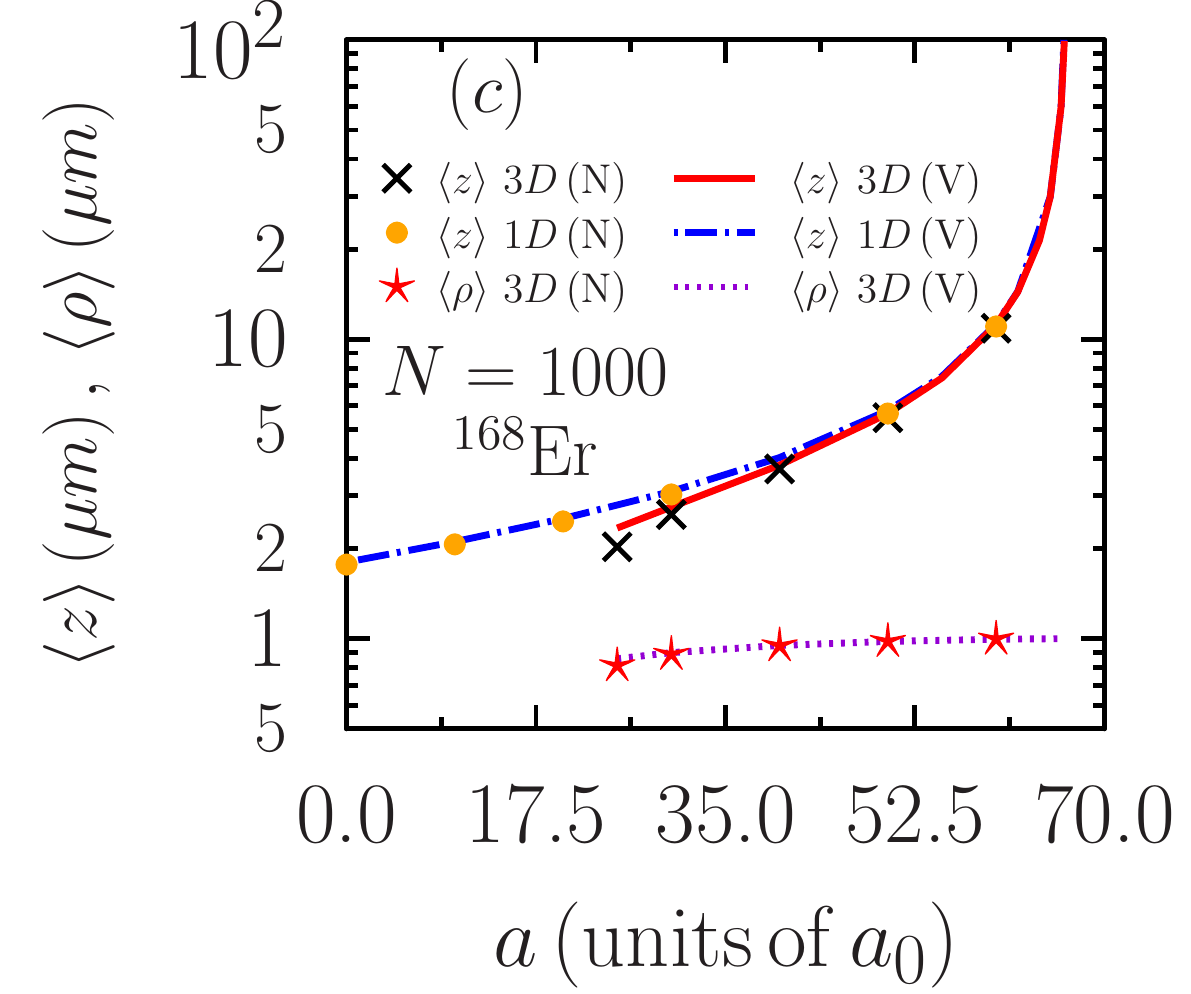}
\includegraphics[width=.49\linewidth]{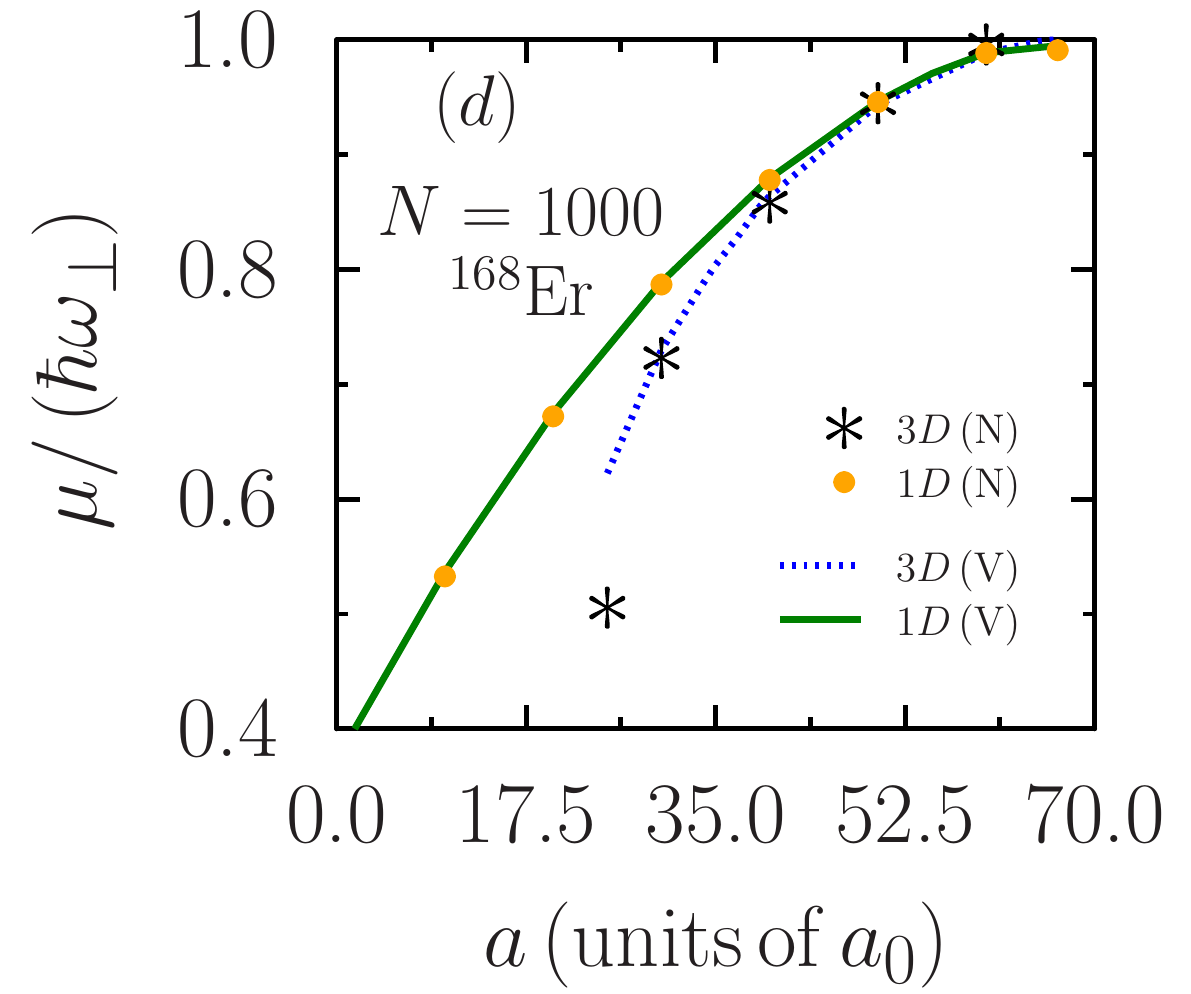}
\includegraphics[width=.49\linewidth]{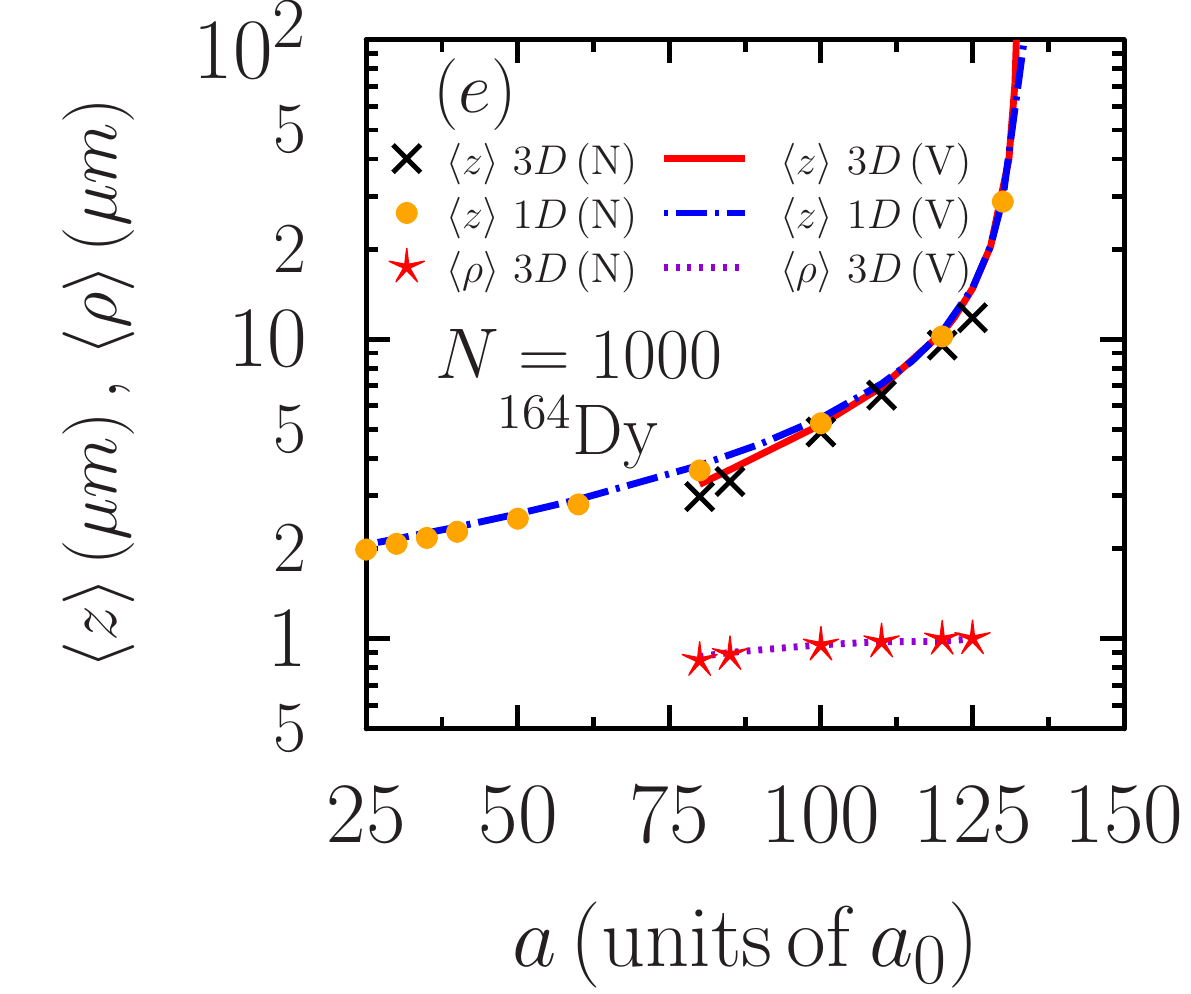}
\includegraphics[width=.49\linewidth]{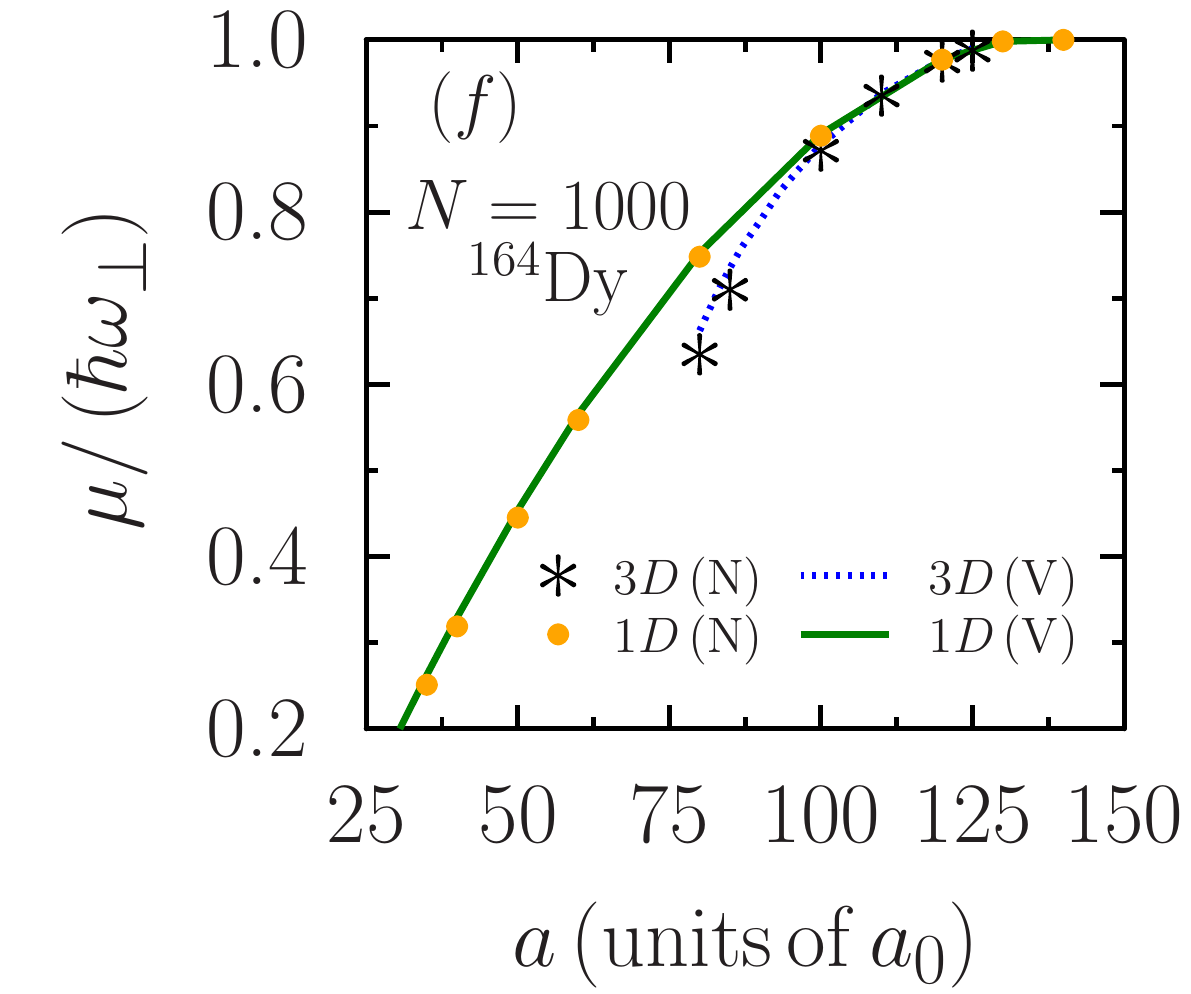}

\caption{Chemical potential $\mu$  and rms size $\left\langle z\right\rangle$ 
versus the scattering length $a$  of a three- and a quasi-1D dipolar 
BEC of 1000 atoms. In a same way it is presented the rms size 
$\left\langle \rho\right\rangle$ for the 3D dipolar condensate. $\left(a\right)$ and $\left(b\right)$
$^{52}\mathrm{Cr}$ $\left(a_{dd}=15a_{0}\right)$. $\left(c\right)$ and $\left(d\right)$ 
$^{168}\mathrm{Er}$
$\left(a_{dd}=66\mathrm{.}6a_{0}\right)$. $\left(e\right)$ and $\left(f\right)$  
$^{164}\mathrm{Dy}$
$\left(a_{dd}=132\mathrm{.}7a_{0}\right)$.
Numerical (N) and variational (V) solution. The oscillator length used is $1\,\mu m$.}
\label{fig3}
\end{center}
\end{figure}

After having established the appropriateness of the 1D reduced equation we plot 
in Figures \ref{fig3}$a$-3$f$ the respective root mean square sizes and the chemical potential
for a 3D and a quasi-1D dipolar condensate.
We did this for three different dipolar BECs 
of  $^{52}$Cr $(a_{dd}=15a_0)$, $^{168}$Er $(a_{dd}=66.6a_0)$ and $^{164}$Dy
$(a_{dd}=132.7a_0)$ atoms in the same regime of 
repulsive atomic interaction $\left(a>0\right)$ each with 1000 atoms. 
The three dipolar BECs in 3D with attractive dipolar interaction 
and repulsive scattering length, should be stable only for a scattering 
length greater than a critical value as we plot in the Figures \ref{fig3}$a$-3$f$. 
This critical value is larger for $^{164}$Dy
atoms compared to that $^{52}$Cr and $^{168}$Er atoms. 

Although the numerical solution of the quasi-1D reduced model
is simpler than the full 3D DGPE, this is still complicated because
of the long-range anisotropic dipolar interaction. The variational approximation 
provides results for the rms size and the chemical potential
in good agreement with the numerical solution of the quasi-1D and full 3D GPE. The
Gaussian ansatz to the reduced model is relatively simple and it could 
be used as an approximate solution.

\begin{figure}
\begin{center}
\newpage\vspace*{-.5cm}
\includegraphics[width=.55\linewidth]{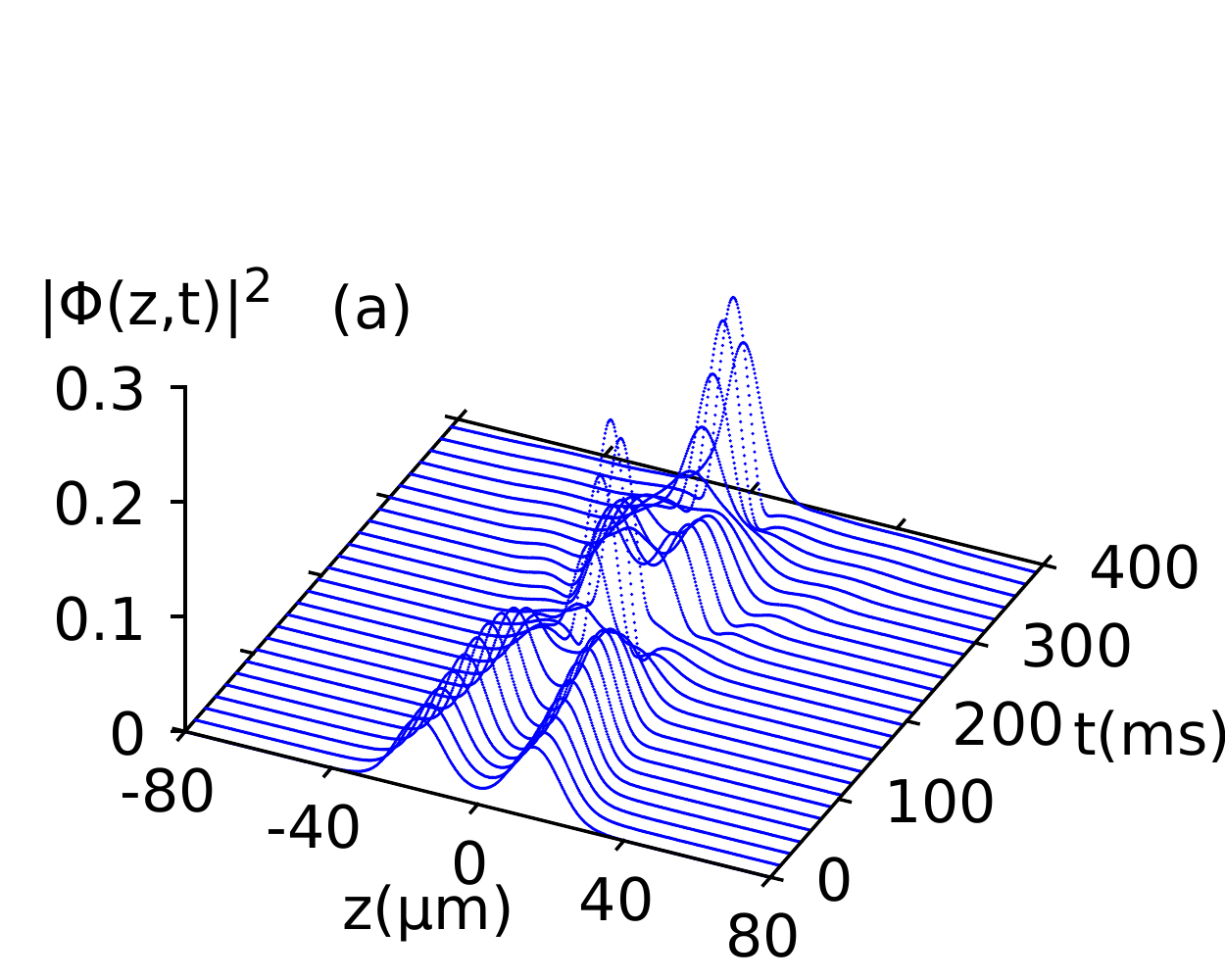}
\includegraphics[width=.43\linewidth]{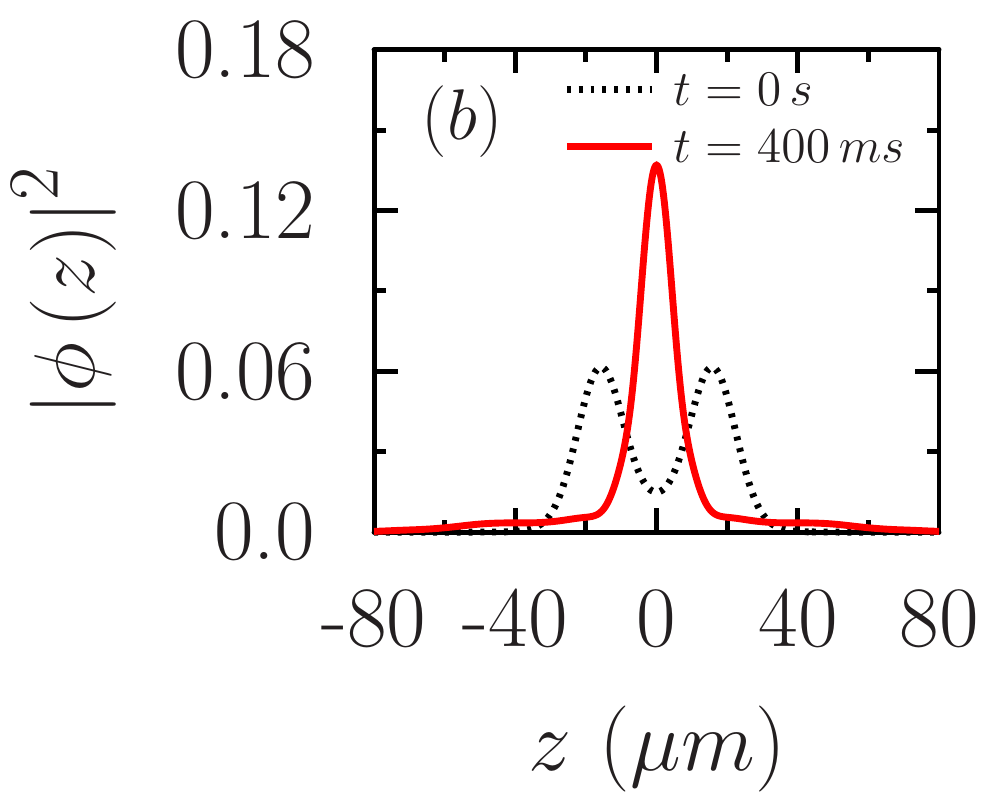}

\includegraphics[width=.55\linewidth]{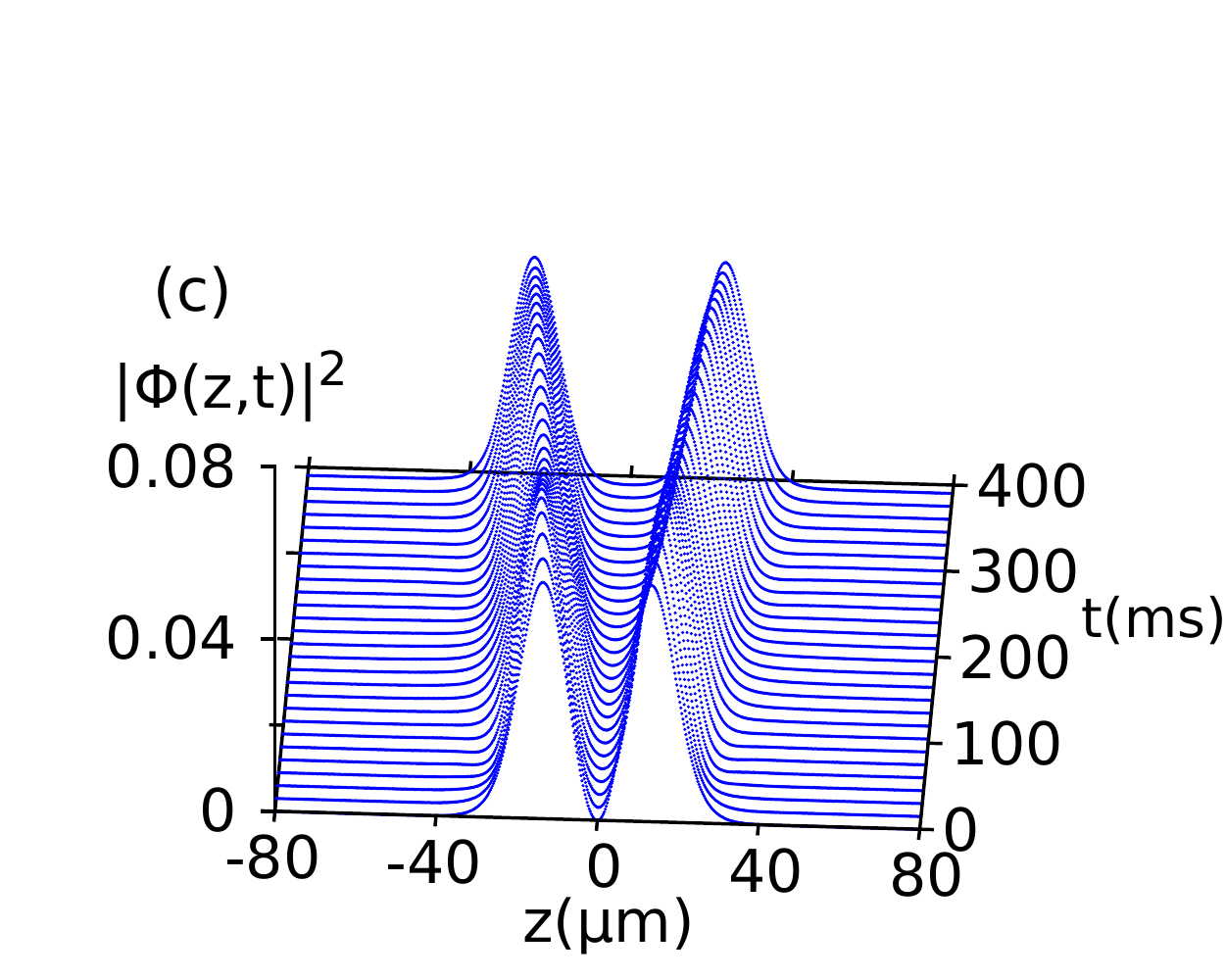}
\includegraphics[width=.43\linewidth]{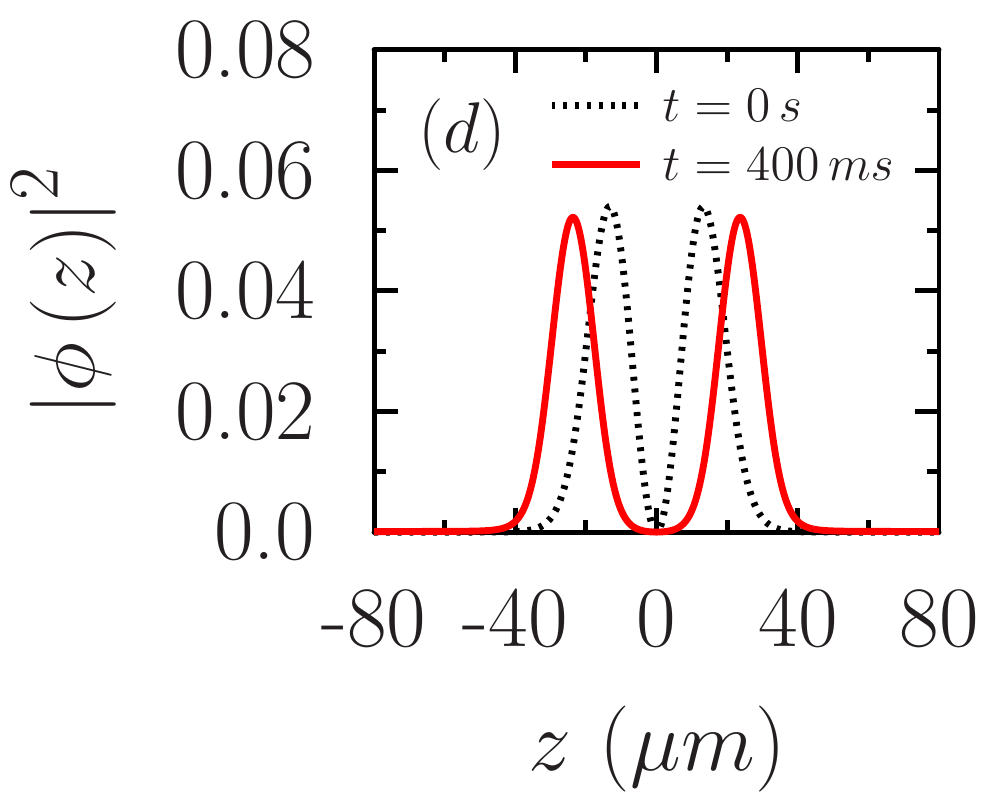}

\caption{The profile of $\left|\phi\left(z,t\right)\right|^{2}$
versus $z$ and $t$ for a train of two bright solitons in a quasi-1D dipolar condensate of 
$^{52}\mathrm{Cr}$ $\left(a_{dd}=15a_{0}\right)$ atoms with atomic scattering length $a=10a_{0}$,
2000 atoms and velocity $v=0$. $(a)$ and $(b)$ correspond to the phase difference $\delta=0$.
$(c)$ and $(d)$ correspond to $\delta=\pi$. The oscillator length used is $1\,\mu m$.}
\label{fig4}
\end{center}
\end{figure}

Now, we investigate the collision between two solitons. 
The numerical solution of the quasi-1D mean-field dipolar GPE
enables the dynamical study of bright solitons. 
To investigate the collision between two bright solitons we apply the following procedure. 
The first step is the creation of one soliton with a number of atoms N=1000,
employing imaginary-time propagation $(t\rightarrow-\rmi t)$ in the Crank-Nicolson method. 
Afterwards two solitons are placed at positions $\pm z_{0}$  with $2z_{0}$
the initial separation between these. 
Then these are advanced by real-time propagation of reduced model (\ref{eq2-red-dip}) 
with N=2000. To introduce the dynamics in the system, the two solitons 
are superposed with a phase difference $\delta$. 
Here we consider the following superposition \cite{bright vortex solitons}
\begin{eqnarray}
\Phi\left(z,t\right)=
\rme^{\rmi \delta}\left|\phi\left(z-z_{0},t\right)\right|+
\left|\phi\left(z+z_{0},t\right)\right|
\end{eqnarray}

Similarly the dynamics can be obtained for a constant velocity $v$ different from zero,
where the soliton placed on the right hand is multiplied by 
a phase factor $\exp\left(\rmi vz\right)$,
while the soliton on the left hand is multiplied by
a phase factor $\exp\left(-\rmi vz\right)$. So the
superposition is given by \cite{Luis Y,Anisotropic 2D sollision}
\begin{eqnarray}
\Phi\left(z,t\right)=
\rme^{\rmi vz}\left|\phi\left(z-z_{0},t\right)\right|+
\rme^{-\rmi vz}\left|\phi\left(z+z_{0},t\right)\right|
\end{eqnarray}

To keep the total number of atoms as a constant
we need to normalize the new wave function of the two solitons
$\Phi\left(z,t\right)$ to 2 because it contains twice the number of atoms
of a single soliton $\phi\left(z,t\right)$.
We present results about the effect of the phase difference
between the solitons  $\delta$, 
for velocities $v=0$ and $v\neq0$ in a train with two equal solitons 
in the three quasi-1D dipolar condensates of $^{52}\mathrm{Cr}$, $^{168}\mathrm{Er}$
and $^{164}\mathrm{Dy}$ atoms. 

In a dipolar condensate of 2000 atoms of $^{52}\mathrm{Cr}$ $\left(a_{dd}=15a_{0}\right)$ 
with scattering length $a=10a_{0}$, velocity $v=0$  and phase difference $\delta=0$ we
plot the time evolution of the train of two such solitons by means of the profile
of $\left|\phi\left(z\right)\right|^{2}$  versus $z$ and $t$. 
In Figures \ref{fig4}$a$ and \ref{fig4}$b$ we show  that 
due to the dipolar attraction, the solitons come close, coalesce and oscillate 
forming a bound soliton molecule around $z=0$. 
We have initial positions at $z=\pm15\mathrm{.}75\,\mu m$ and an interval of time $400\, ms$.
For a phase difference $\delta=\pi$
in an interval of time $400\, ms$ the two solitons repel and stay away from each other,
Figure \ref{fig4}$c$ and \ref{fig4}$d$.
These moved from positions $z=\pm13\mathrm{.}25\,\mu m$ to $z=\pm23\mathrm{.}75\,\mu m$.
In this short time because of the long-range character of the dipolar interaction the solitons
still remain interacting. The final solitons are different to the initial ones. 
Our numerical results show that we need increasing the time evolution by three or four times
to get the final solitons very similar to the initial ones.

\begin{figure}
\begin{center}
\newpage\vspace*{-.5cm}
\includegraphics[width=.55\linewidth]{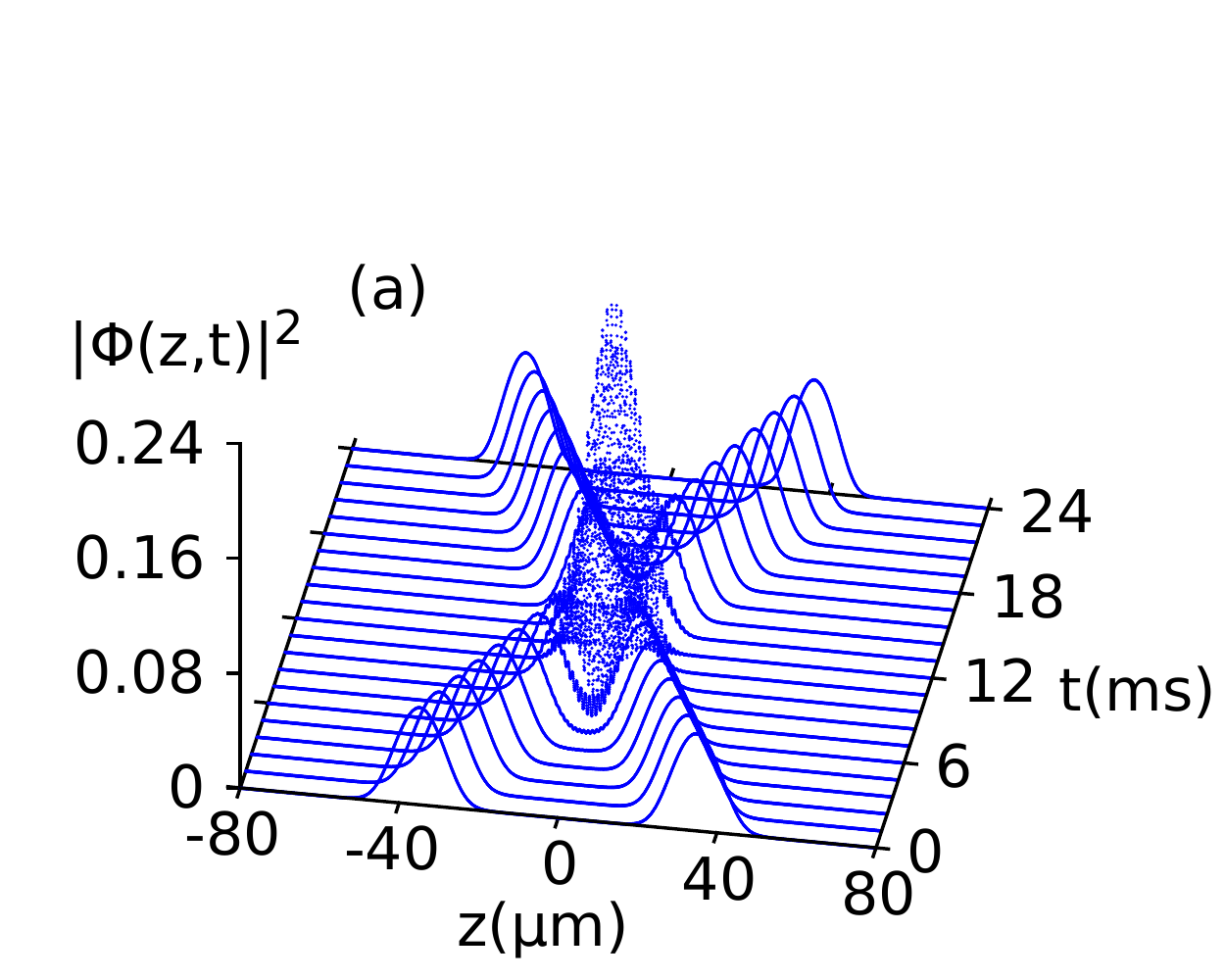}
\includegraphics[width=.43\linewidth]{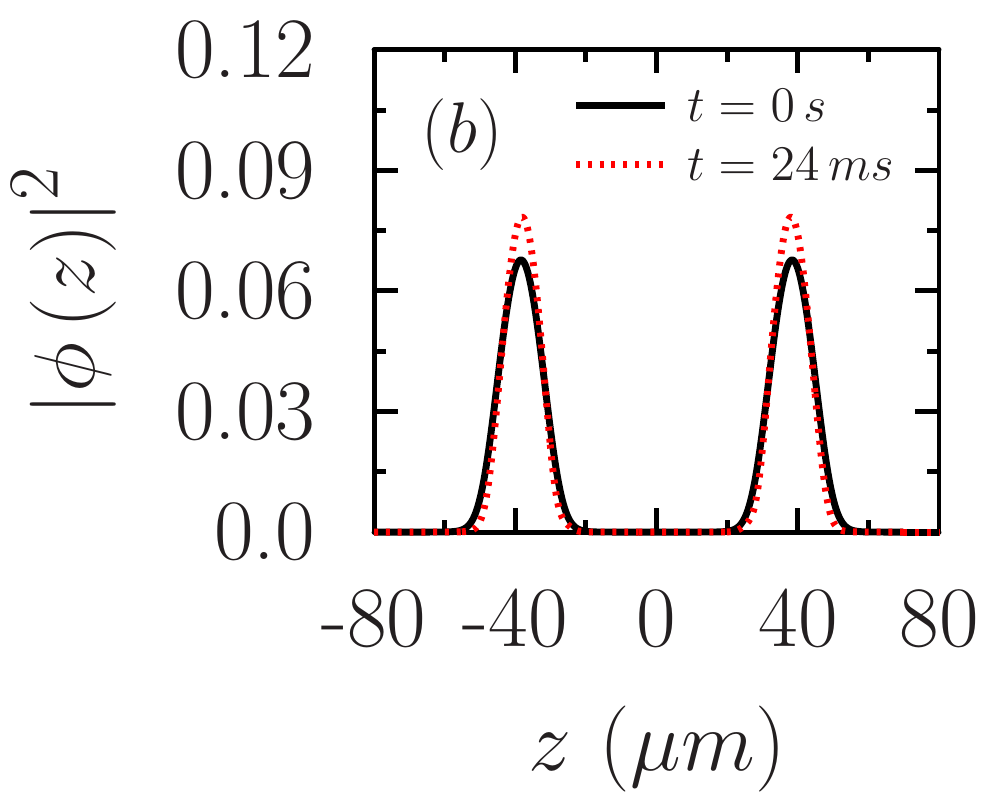}

\caption{$(a)$ and $(b)$. 
The profile of  $\left|\phi\left(z,t\right)\right|^{2}$
versus $z$ and $t$ for a train of two bright solitons in a quasi-1D dipolar condensate of
$^{168}\mathrm{Er}$ $\left(a_{dd}=66\mathrm{.}6a_{0}\right)$  atoms with atomic scattering 
length $a=50a_{0}$, 2000 atoms and velocity $v=3\,\mathrm{mm\, s^{-1}}$.
The oscillator length used is $1\,\mu m$.}
\label{fig5}
\end{center}
\end{figure}

In Figures \ref{fig5}$a$ and  \ref{fig5}$b$
we illustrate the collision between two bright solitons 
in a dipolar condensate of 2000 atoms of $^{168}\mathrm{Er}$
$\left(a_{dd}=66\mathrm{.}6a_{0}\right)$ with scattering length $a=50a_{0}$. 
The collision is insensitive to the initial phase difference $\delta=0$  or $\delta=\pi$
when the velocity is $v=3\,\mathrm{mm\, s^{-1}}$. 
The solitons come towards each other and interact at $z=0$. 
Then these are separated and continue practically unchanged. 
The solitons are placed at $z=\pm38\mathrm{.}4\,\mu m$  at $t=0$ and each of these is
advanced with a constant velocity $v=3\,\mathrm{mm\, s^{-1}}$ towards centre $z=0$.
The real-time simulation of quasi-1D model (\ref{eq2-red-dip}) is terminated when the solitons 
reach approximately $z=\mp38\mathrm{.}4\,\mu m$  at time about $t=24\, ms$. 
The final solitons are not equal to the initial ones. 
So the interaction is quasi-elastic and the final result is two quasi-solitons.

\begin{figure}
\begin{center}
\newpage\vspace*{-.5cm}
\includegraphics[width=.55\linewidth]{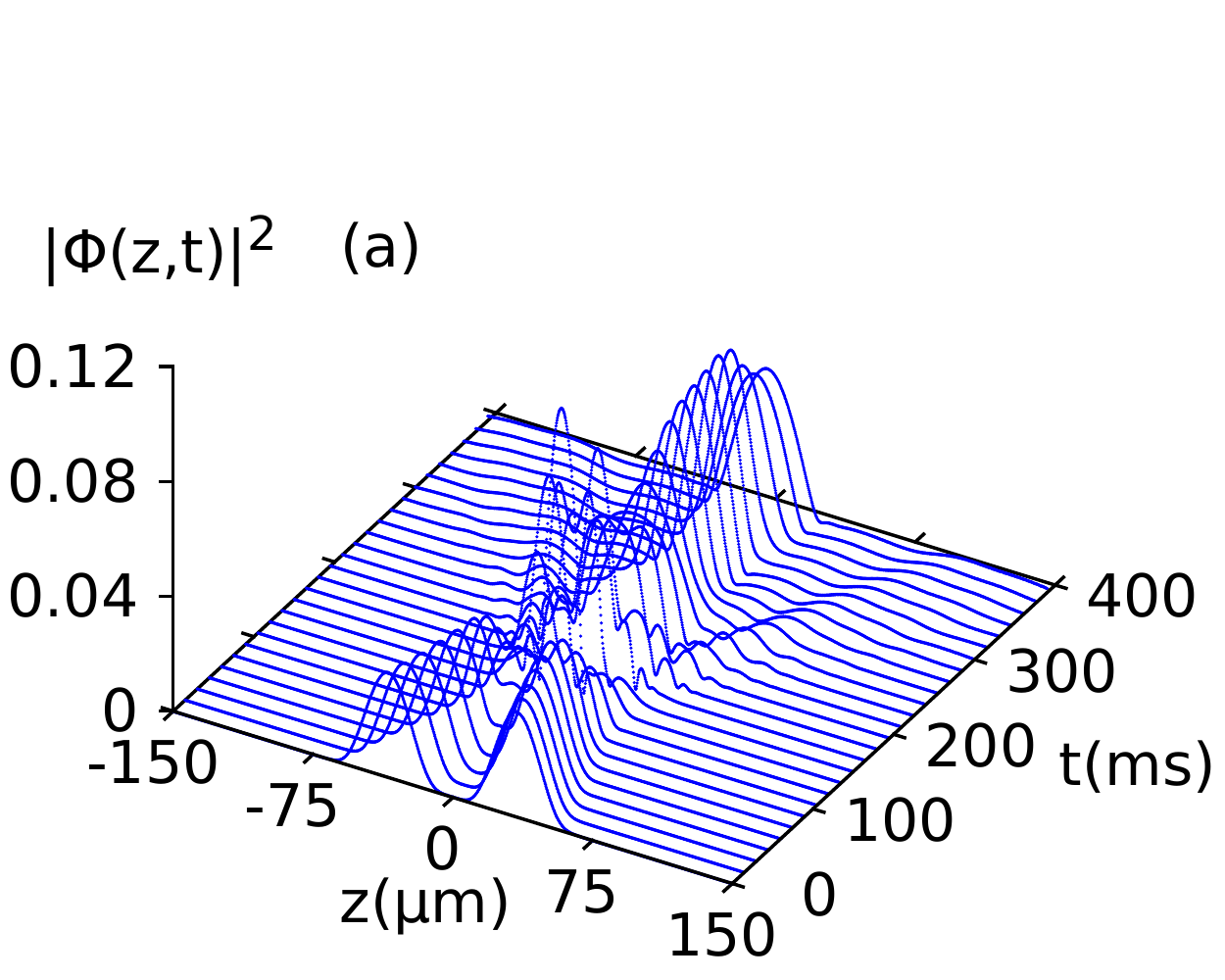}
\includegraphics[width=.43\linewidth]{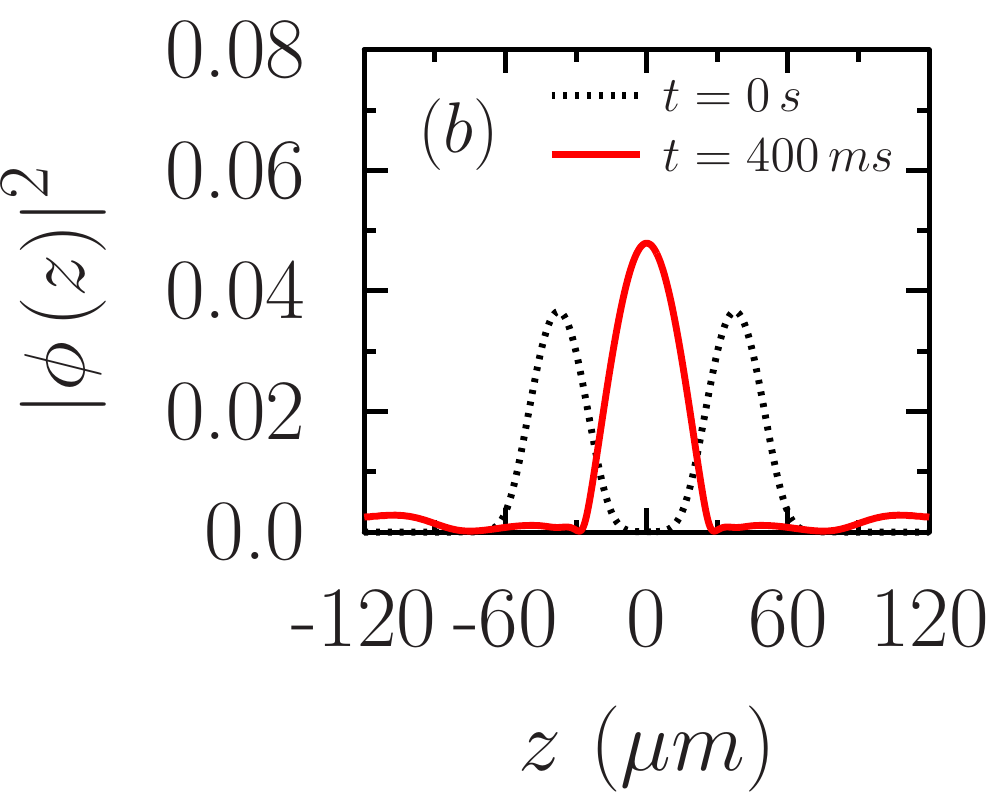}

\includegraphics[width=.55\linewidth]{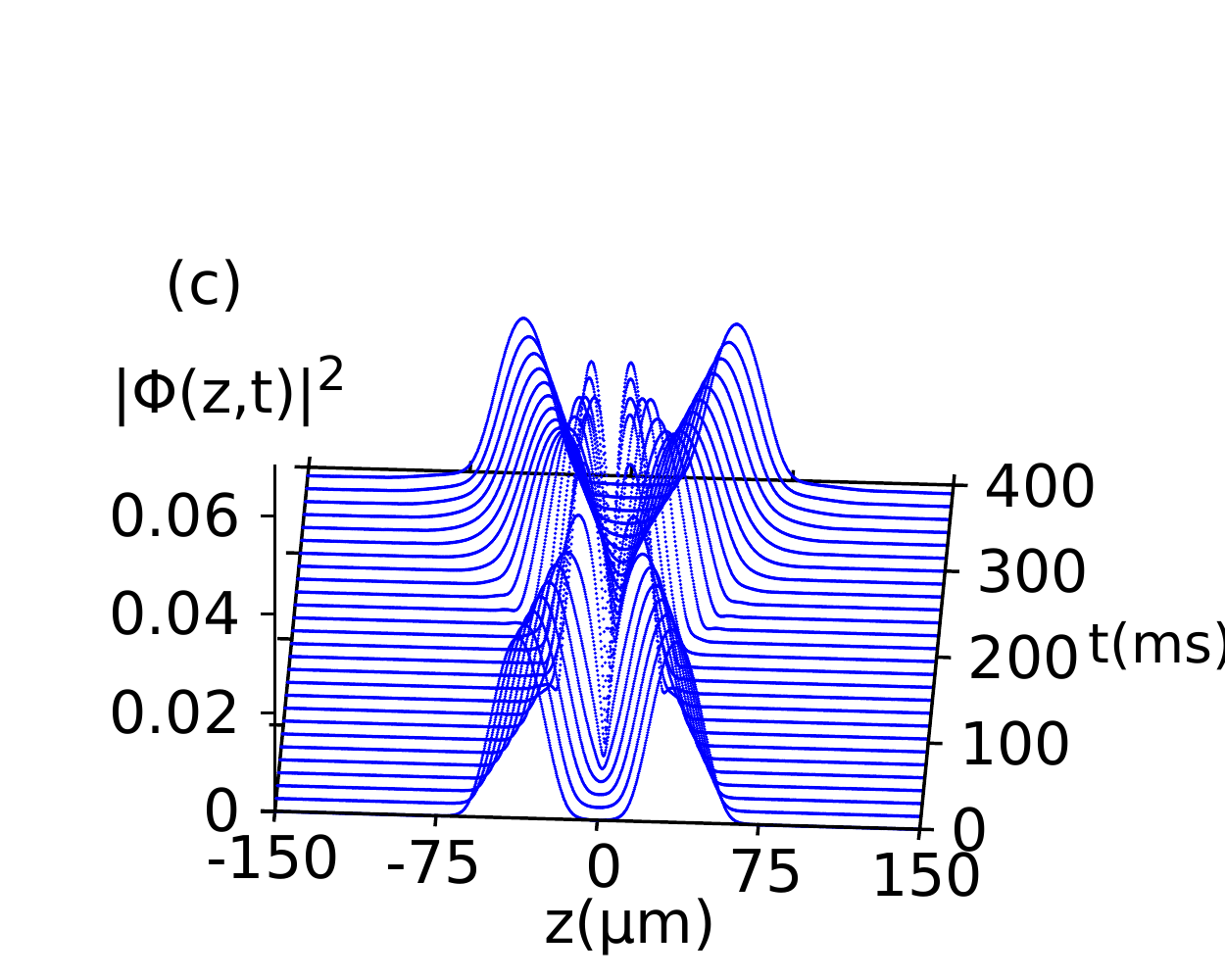}
\includegraphics[width=.43\linewidth]{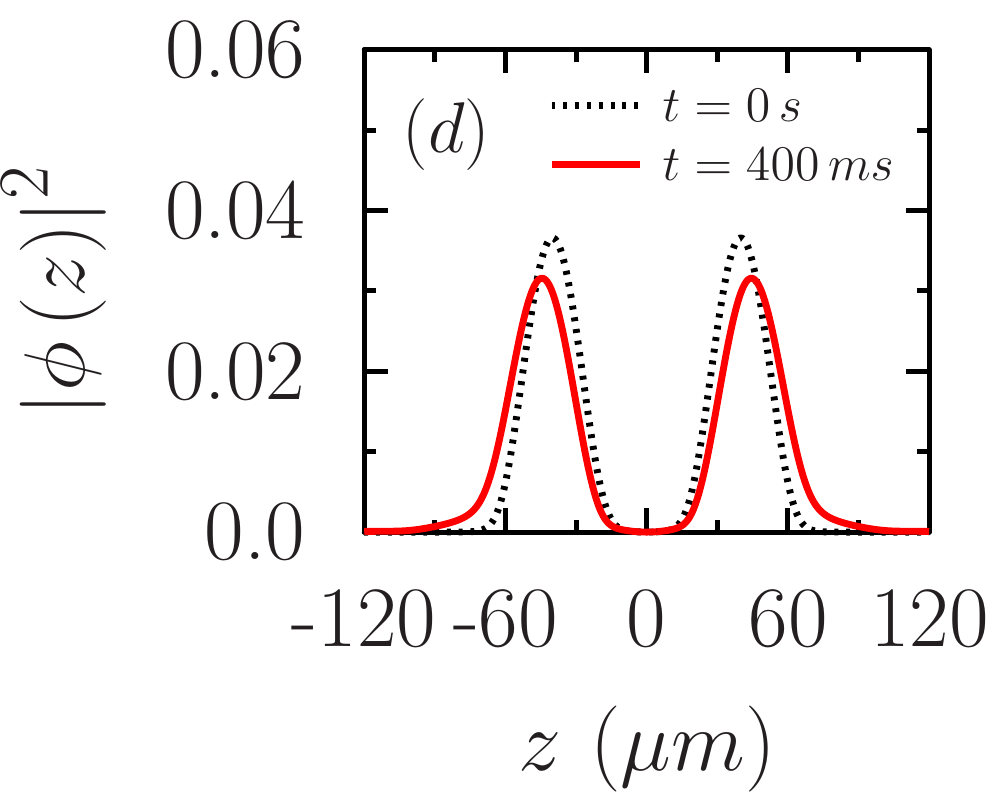}

\caption{The profile of  $\left|\phi\left(z,t\right)\right|^{2}$
versus $z$ and $t$ of two bright solitons in a quasi-1D dipolar condensate of
$^{164}\mathrm{Dy}$ $\left(a_{dd}=132\mathrm{.}7a_{0}\right)$ atoms with atomic scattering 
length $a=120a_{0}$, 2000 atoms and velocity $v=0\mathrm{.}2\,\mathrm{mm\, s^{-1}}$.
$(a)$ and $(b)$ correspond to $\delta=0$. $(c)$ and $(d)$ correspond
to $\delta=\pi$. The oscillator length used is $1\,\mu m$.}
\label{fig6}
\end{center}
\end{figure}

Finally, we study the effect of the phase difference $\delta=0$ and $\delta=\pi$
for the collision of two bright solitons in a dipolar condensate of 2000
atoms of $^{164}\mathrm{Dy}$  $\left(a_{dd}=132\mathrm{.}7a_{0}\right)$, 
with scattering length $a=120a_{0}$ and velocity $v=0\mathrm{.}2\,\mathrm{mm\, s^{-1}}$.
As we show in the Figures \ref{fig6}$a$ - \ref{fig6}$d$, the collision is sensitive to the initial phase difference. 
In an interval of time $400\, ms$  with initial positions $z=\pm37\mathrm{.}5\,\mu m$
and $\delta=0$ the solitons come towards each other interact and the dipolar attraction 
along with the attractive phase difference allow the solitons together, thus
forming a bound soliton molecule around $z=0$, Figures \ref{fig6}$a$ and \ref{fig6}$b$.
This happens in a similar way as when the velocity is zero in the 
Figures \ref{fig4}$a$ and \ref{fig4}$b$. 
For a phase difference $\delta=\pi$ in an interval of time $400\, ms$
the two solitons come towards each other and interact but (unlike when the velocity is zero 
in Figures \ref{fig4}$c$ and \ref{fig4}$d$)
do not coalesce because of the repulsive effect of the phase difference,
Figures \ref{fig6}$c$ and \ref{fig6}$d$. So these repel and stay away from each other. 
The solitons were placed at initial positions $z=\pm40\,\mu m$
and these reach the final state at $z=\pm44\mathrm{.}5\,\mu m$. At $400\, ms$
the long-range dipolar interaction still maintains
the interaction between the solitons and their shapes are not the same of
initial ones. To recovering the initial solitons we need increasing the system time evolution.

\section{Summary and discussion}\label{sec4}

To find bright solitons numerically in a 3D dipolar condensate 
using the mean-field non-local Gross-Pitaevskii equation is a complicated issue because of the
anisotropic long-range character of the dipolar interaction. 
Nevertheless, the reduced quasi-1D equation provides an alternative to the complete 3D equation. 
We show the existence of bright solitons in a quasi-1D reduced model of a
dipolar BEC as result of balance between the 
repulsive short-range contact interaction
and the anisotropic long-range dipolar attraction.
We show this with plots of the rms size and the chemical potential for three
different condensates of $^{52}$Cr, $^{168}$Er and $^{164}$Dy atoms.  
Our findings show that the bright solitons exist for any value of the scattering length 
less than the dipolar strength. There is no collapse in this reduced model 
in contrast to the 3D full equation.
The Gaussian variational approximation in the quasi-1D model is relatively 
simple and it provides results for
the stationary cigar-shaped DBEC in good agreement with the numerical solution of the GPE. 
The collision between two bright solitons at small velocities shows that 
for a phase difference $\delta=0$ we have a bound soliton molecule due to the dipolar attraction.
When the phase difference is $\delta=\pi$ the solitons repel and move away.
However, the long-range character of the dipolar contribution still maintains the interaction
between the solitons.
At high velocities the collision is quasi-elastic and it is 
independent of the initial phase difference $\delta$.

\ack
I thank CAPES (Brazil) for partial support. 
I thank Sadhan Adhikari for useful comments on a preliminary version of the paper.


\section*{References}
\begin{thebibliography}{99}


\bibitem{solit-1}P\'erez-Garc\'ia V M, Michinel H and Herrero H 1998 {\it Phys. Rev. A }
{\bf 57} 3837

\bibitem{solit-Li-1}Strecker K E, Partridge G B, Truscott A G
and Hulet R G 2002 {\it Nature} {\bf 417},  150

\bibitem{solit-Li-2}Khaykovich L {\it et al.} 2002 {\it Science} {\bf 256} 1290

\bibitem{solit-Rb}Cornish S L, Thompson S T and Wieman C E 2006 {\it Phys. Rev. Lett.}
{\bf 96} 170401


\bibitem{Cr-1} Lahaye T {\it et al.} 2009 {\it Rep. Prog. Phys.} {\bf 72} 126401
\bibitem{Cr-2} Lahaye T {\it et al.} 2007 {\it Nature} {\bf 448} 672 
\bibitem{Cr-3} Stuhler J {\it et al.} 2005 {\it Phys. Rev. Lett.} {\bf 95} 150406
\bibitem{Cr-4} G\'oral K, Rzazewski K and Pfau T 2000 {\it Phys. Rev. A} {\bf 61} 051601
\bibitem{Cr-5} Koch T {\it et al.}, 2008 {\it Nature Phys.} {\bf 4} 218
\bibitem{Cr-6} Griesmaier A {\it et al.}, 2006 {\it Phys. Rev. Lett.} {\bf 97} 250402


\bibitem{Dy-1} Lu M, Youn S H and Lev B L 2010 {\it Phys. Rev. Lett.} {\bf 104},
063001 

 McClelland J J and Hanssen J L 2006 {\it Phys. Rev. Lett.} {\bf 96}, 143005 
 
 Youn S H, Lu M W, Ray U and Lev B V 2010 {\it Phys. Rev. A} {\bf 82}, 043425


\bibitem{Er-1} Aikawa K, Frisch A, Mark M, Baier S, Rietzler A, Grimm R
and Ferlaino F 2012 {\it Phys. Rev. Lett.} {\bf 108} 210401


\bibitem{Electric}  Deiglmayr J, Grochola A, Repp M, M\"ortlbauer K, 
Gl\"uck C, Lange J, Dulieu O, Wester R and Weidem\"uller M 2008 
{\it Phys. Rev. Lett.} {\bf 101} 133004 

 de Miranda M H G {\it et al.} 2011  {\it Nature} Phys. {\bf 7} 502


\bibitem{Two Dim Bright} Pedri P and Santos L 2005 {\it Phys. Rev. Lett.}
{\bf 95} 200404 


\bibitem{Anisotropic Solitons} Tikhonenkov I, Malomed B and Vardi A 2008
{\it Phys. Rev. Lett.} {\bf 100} 090406 

\bibitem{Anisotropic 2D sollision}Eichler R, Zajec D, K\"{o}berle P, 
Main J and Wunner G 2012 {\it Phys. Rev. A}  {\bf 86}, 053611 

\bibitem{Experimental anisotropic 2D-solitons} K\"{o}berle P, Zajec D, Wunner G
and Malomed B A 2012 {\it Phys. Rev. A} {\bf 85} 023630


\bibitem{Solitons induced} Li Y, Liu J, Pang W and Malomed B A
2013 {\it Phys. Rev. A} {\bf 88} 053630



\bibitem{Cuevas} Cuevas J, Malomed B, Kevrekidis P and 
Frantzeskakis D 2009 {\it Phys. Rev. A} {\bf 79} 053608


\bibitem{Luis Y} Young-S L E, Muruganandam P and Adhikari S K 2011 
{\it J. Phys. B: At. Mol. Opt. Phys.} {\bf 44} 101001


\bibitem{Dim-reduc} Muruganandam P and Adhikari S K 2012 {\it Laser Physics} 
{\bf 22} 813-820


\bibitem{libro Pethick} Pethick C J and Smith H 2008 
{\it Bose-Einstein Condensation in Dilute Gases}
Second edition (New York: Cambridge University Press) p 183 

\bibitem{art Dalfovo}Dalfovo F, Giorgini S, Pitaevskii L,
Stringari S 1999 {\it Rev. Mod. Phys.} {\bf 71}, 463


\bibitem{Effective wave equations} Salasnich L, Parola A, and Reatto
L 2002 {\it Phys. Rev. A} {\bf 65}, 043614-1

\bibitem{Effective mean-field equations for} Mu\~{n}oz A and Delgado V 2008
{\it Phys. Rev. A} {\bf 77}, 013617

\bibitem{reduction} Buitrago C and Adhikari S J 2009
{\it Phys. B: At. Mol. Opt. Phys.} {\bf 42} 215306 


\bibitem{97 Exact hydrodynamics dbec} Eberlein C, Giovanazzi S, 
and O'Dell D 2005 {\it Phys. Rev. A} {\bf 71}, 033618 

\bibitem{funcion f(k)} Glaum K and Pelster A 2007 {\it Phys. Rev. A} {\bf 76}, 023604


\bibitem{Adhikari comput}Muruganandam P and Adhikari S K  2009
{\it Computer Physics Communications} {\bf 180} 1888–1912


\bibitem{Mean-field regime of trapped DBEC}Cai Y, Rosenkranz M, Lei Z, and Bao W 2010
{\it Phys. Rev. A} {\bf 82} 043623 


\bibitem{Trans Fourier} Salomon C, Shlyapnikov G, and Cugliandolo L 2013 
{\it Many-Body Physics with Ultracold Gases:
Lecture Notes of the Les Houches 2010}
(Great Britain: Oxford University Press) p 235

\bibitem{Fast FT} G\'oral K and Santos L 2002 {\it Phys. Rev. A} {\bf 66} 023613


\bibitem{bright vortex solitons} Adhikari S 2003 {\it New Journal of Physics} 
v. {\bf 5}, p. 137.1-.13

\end{thebibliography}
\end{document}